\documentstyle[aps,preprint,eqsecnum,floats,epsf]{revtex}
\tighten
\draft
\begin{document}

\preprint{} \title{A Generalized Jaynes-Cummings Hamiltonian and
Supersymmetric Shape-Invariance}

\author{A.~N.~F. Aleixo\thanks{Permanent address: Instituto de
F\'{\i}sica,  Universidade Federal  do Rio de Janeiro, RJ -  Brazil. 
Electronic address: {\tt aleixo@nucth.physics.wisc.edu}}} 
\address{Department of Physics,
University of Wisconsin;\\ 
Madison, Wisconsin 53706 USA}

\author{A.~B. Balantekin\thanks{Permanent address: Department of
Physics, University of Wisconsin,  Madison, Wisconsin 53706 USA. 
Electronic address: {\tt baha@nucth.physics.wisc.edu}}} 
\address{Max-Planck-Institut f\"ur Kernphysik,
Postfach 103980, D-69029 Heidelberg, Germany} 

\author{M.~A. C\^andido Ribeiro\thanks{Electronic address: {\tt
macr@df.ibilce.unesp.br}},} 
\address{Departamento de F\'{\i}sica -
Instituto de Bioci\^encias, Letras e Ci\^encias Exatas\\ UNESP, S\~ao
Jos\'e do Rio Preto, SP - Brazil}

\maketitle

\begin{abstract}
  A class of shape-invariant bound-state problems which represent
  two-level systems are introduced. It is shown that the
  coupled-channel Hamiltonians obtained correspond to the
  generalization of the Jaynes-Cummings Hamiltonian.   
\end{abstract}


\newpage
\section{Introduction}

Supersymmetric quantum mechanics \cite{ref1,ref2} deals with pairs of
Hamiltonians which have the same energy spectra, but different
eigenstates. A number of such pairs of Hamiltonians share an
integrability condition called shape invariance \cite{ref3}. Although
not all exactly-solvable problems are shape-invariant \cite{ref4},
shape invariance, especially in its algebraic formulation
\cite{ref5,ref6,ref7}, is a powerful technique to study
exactly-solvable systems. 

Supersymmetric quantum mechanics is generally studied in the context
of one-dimensional systems. The partner Hamiltonians
\begin{mathletters}
\label{eqh12}
\begin{eqnarray}
\hat H_1 &=& \hat A^\dagger\hat A  \\ \hat H_2 &=& \hat A\hat
A^\dagger \,,
\end{eqnarray}
\end{mathletters}
are most readily written in terms of one-dimensional operators
\begin{mathletters}
\label{eqaa+}
\begin{eqnarray}
\hat A &\equiv& W(x) + \frac{i}{\sqrt{2m}}\hat p\,, \\ \hat A^\dagger
&\equiv& W(x) - \frac{i}{\sqrt{2m}}\hat p\,,
\end{eqnarray}
\end{mathletters}
where $W(x)$ is the superpotential. Attempts were made to generalize
supersymmetric quantum mechanics and the concept of shape-invariance
beyond one-dimensional and spherically-symmetric three-dimensional
problems. These include non-central \cite{ref8}, non-local
\cite{ref9}, and periodic \cite{ref10} potentials; a three-body
problem in one-dimension \cite{ref11} with a three-body force
\cite{ref12}; N-body problem \cite{ref13}; and coupled-channel
problems \cite{ref14,ref15}. It is not easy to find exact solutions to
these problems. For example, in the coupled-channel case a general
shape-invariance is only possible in the limit where the
superpotential is separable \cite{ref15} which corresponds to the
well-known sudden approximation in the coupled-channel problem
\cite{ref16}. Our goal in this article is to introduce a class of
shape-invariant coupled-channel problems which correspond to the
generalization of the Jaynes-Cummings Hamiltonian \cite{ref17}.

\section{Shape Invariance}

The Hamiltonian $\hat H_1$ of Eq.~(\ref{eqh12}) is called
shape-invariant if the condition
\begin{equation}
\hat A(a_1) \hat A^\dagger(a_1) =\hat A^\dagger (a_2)  \hat A(a_2) +
R(a_1) \,,
\label{eqsi}
\end{equation}
is satisfied \cite{ref3}. In this equation $a_1$ and $a_2$ represent
parameters of the Hamiltonian. The parameter $a_2$ is a function of
$a_1$ and the remainder $R(a_1)$ is independent of the dynamical
variables such as position and momentum. As it is written the
condition of Eq.~(\ref{eqsi}) does not require the Hamiltonian to be
one-dimensional, and one does not need to choose the ansatz of
Eq.~(\ref{eqaa+}). In the cases studied so far the parameters $a_1$
and $a_2$ are either related by a translation \cite{ref4,ref18} or a
scaling \cite{ref19}. Introducing the similarity transformation that
replaces $a_1$ with $a_2$ in a given operator 
\begin{equation}
\hat T(a_1)\, \hat O(a_1)\, \hat T^\dagger(a_1) = \hat O(a_2)
\label{eqsio}
\end{equation}
and the operators
\begin{equation}
\hat B_+ =  \hat A^\dagger(a_1)\hat T(a_1)
\label{eqba}
\end{equation}
\begin{equation}
\hat B_- =\hat B_+^\dagger =  \hat T^\dagger(a_1)\hat A(a_1)\,,
\label{eqbe}
\end{equation}
the Hamiltonians of Eq.~(\ref{eqh12}) take the forms
\begin{equation}
\hat H_1 =\hat B_+\hat B_-\,.
\label{eqhb1}
\end{equation}
and
\begin{equation}
\hat H_2 = \hat T\hat B_-\hat B_+\hat T^\dagger\,.
\label{eqhb2}
\end{equation}
Using Eq.~(\ref{eqsi}) one can also easily prove the commutation
relation \cite{ref5}
\begin{equation}
[\hat B_-,\hat B_+] =  \hat T^\dagger(a_1)R(a_1)\hat T(a_1)  \equiv
R(a_0)\,,
\label{eqcb1}
\end{equation}
where we used the identity
\begin{equation}
R(a_n) = {\hat T}(a_1)\,R(a_{n-1})\,{\hat T}^\dagger (a_1)\,,
\label{eqran}
\end{equation}
valid for any $n$. The ground state of the Hamiltonian $\hat H_1$ 
satisfies the condition
\begin{equation}
\hat A\,\mid \psi_0\rangle = 0 = \hat B_-\,\mid\psi_0\rangle\,.
\label{eqaps0}
\end{equation}
The $n$-th excited state of $\hat H_1$ is given by 
\begin{equation}
\mid \psi_n\rangle \sim \left( \hat B_+ \right)^n \mid\psi_0\rangle
\label{eqpsn}
\end{equation}
with the eigenvalue
\begin{equation}
\varepsilon_n =  \sum_{k=1}^n R(a_k)\,.
\label{eqen}
\end{equation}
Note that the eigenstate of Eq.~(\ref{eqpsn}) needs  to be suitably
normalized. We discuss the normalization of this state in the next
section. 

\section{Generalization of the Jaynes-Cummings Hamiltonian}

To generalize the Jaynes-Cummings Hamiltonian to general
shape-invariant systems we introduce the operator
\begin{equation}
\hat S = \sigma_+\hat A + \sigma_-\hat A^\dagger\,,
\label{eqso}
\end{equation}
where 
\begin{equation}
\sigma_\pm = {1\over 2}\left( \sigma_1\pm i\sigma_2\right)\,,
\label{eqsig}
\end{equation}
with $\sigma_i$, with $i=1,\,2,\, {\rm and}\,\, 3$, being the Pauli
matrices and the operators $\hat A$ and $\hat A^\dagger$ satisfy the
shape invariance condition of Eq.~(\ref{eqsi}). We search for the
eigenstates of $\hat S$. It is more convenient to work with the square
of this operator, which can be written as
\begin{equation}
\hat S^2 = \left[ \matrix{\hat T & 0\cr 0 & \pm 1 \cr}\right] \left[
\matrix{\hat B_-\hat B_+ & 0\cr 0 & \hat B_+\hat B_- \cr}\right]
\left[ \matrix{\hat T^\dagger & 0\cr 0 & \pm 1 \cr}\right]\,.
\label{eqso2}
\end{equation}
Note the freedom of sign choice in this equation, which results in two
possible decompositions of $\hat S^2$. 

We next introduce the states
\begin{equation}
\mid \Psi\rangle_\pm =  \left[ \matrix{\hat T & 0\cr 0 & \pm 1
\cr}\right]  \left[ \matrix{\mid m\rangle\cr \mid n\rangle \cr}\right]
\label{eqps+-}
\end{equation}
where $\mid m\rangle$ and $\mid n\rangle$ are the abbreviated notation
for the states $\mid \psi_n\rangle$ and $\mid \psi_m\rangle$ of
Eq.~(\ref{eqpsn}). Using Eqs.~(\ref{eqcb1}), (\ref{eqso2}) and
(\ref{eqps+-}) and the fact that the operator $\hat T$ is unitary one
gets
\begin{eqnarray}
\hat S^2\mid \Psi\rangle_\pm &=&  \left[ \matrix{\hat T & 0\cr 0 & \pm
1 \cr}\right] \left[ \matrix{\hat B_+\hat B_- + R(a_0) & 0\cr 0 & \hat
B_+\hat B_- \cr}\right]\left[ \matrix{\mid m\rangle\cr \mid n\rangle
\cr}\right]  \nonumber\\ &=& \left[ \matrix{\hat T & 0\cr 0 & \pm 1
\cr}\right] \left[ \matrix{\varepsilon_m + R(a_0) & 0\cr 0 &
\varepsilon_n \cr}\right]\left[ \matrix{\mid m\rangle\cr \mid n\rangle
\cr}\right]\,. 
\label{eqmsop2}
\end{eqnarray}
Using Eqs.~(\ref{eqran}) and (\ref{eqen}) one can write
\begin{eqnarray}
\hat T\left[ \varepsilon_m + R(a_0)\right]\hat T^\dagger &=&  \hat
T\left[ R(a_1) + R(a_2) + \cdots + R(a_m) + R(a_0)\right]\hat
T^\dagger \nonumber\\ &=&  R(a_2) + R(a_3) + \cdots + R(a_{m+1}) +
R(a_1) = \varepsilon_{m+1}\,.
\label{eqtemt}
\end{eqnarray}
Hence the states 
\begin{equation}
\mid \Psi_m\rangle_\pm =  \frac{1}{\sqrt{2}}\left[ \matrix{\hat T &
0\cr 0 & \pm 1 \cr}\right] \left[ \matrix{\mid m\rangle\cr \mid
m+1\rangle \cr}\right], \> m=0,1,2, \cdots \label{eqpsm+-}
\end{equation}
are the normalized eigenstates of the operator $\hat S^2$
\begin{equation}
\hat S^2\mid \Psi_m\rangle_\pm = \varepsilon_{m+1}\mid
\Psi_m\rangle_\pm\,.
\label{eqs2psm}
\end{equation}
One can also calculate the action of the operator $\hat S$ on this
state
\begin{equation}
\hat S\mid \Psi_m\rangle_\pm = \frac{1}{\sqrt{2}} 
\left[ \matrix{\pm\hat T\hat B_-\mid
m+1\rangle\cr \hat B_+\mid m\rangle \cr}\right]\,.
\label{eqspsm}
\end{equation}
Introducing the operator \cite{ref7}
\begin{equation}
\hat Q^\dagger = \left(\hat B_+\hat B_-\right)^{-1/2}\hat B_+
\label{eqq+}
\end{equation}
one can write the normalized eigenstate of $\hat H_1$ as 
\begin{equation}
\mid m\rangle = \left( \hat Q^\dagger\right)^m\mid 0\rangle\,.
\label{eqsmq}
\end{equation}
Using Eqs.~(\ref{eqq+}) and (\ref{eqsmq}) one gets
\begin{equation}
\hat B_+\mid m\rangle = \sqrt{\varepsilon_{m+1}}\mid m+1\rangle\,. 
\label{eqb+m}
\end{equation}
Similarly
\begin{eqnarray}
\hat T\hat B_-\mid m+1\rangle &=& \hat T\hat B_-{1\over\sqrt{\hat
B_+\hat B_-}}\hat B_+\mid m\rangle\nonumber\\ &=& \hat T\sqrt{\hat
B_-\hat B_+}\mid m\rangle\nonumber\\ &=& \hat T\sqrt{\varepsilon_m +
R(a_0)}\mid m\rangle\nonumber\\ &=& \sqrt{\varepsilon_{m+1}}\,\hat
T\mid m\rangle\,.
\label{eqtbm+1}
\end{eqnarray}
Using Eqs.~(\ref{eqb+m}) and (\ref{eqtbm+1}), Eq.~(\ref{eqspsm}) takes
the form
\begin{eqnarray}
\hat S\mid \Psi_m\rangle_\pm &=& \frac{1}{\sqrt{2}}
\sqrt{\varepsilon_{m+1}} \left[
\matrix{\pm\hat T\mid m\rangle\cr \mid m+1\rangle \cr}\right]
\nonumber\\ &=& \pm\sqrt{\varepsilon_{m+1}}\mid \Psi_m\rangle_\pm\,.
\label{eqspsifi}
\end{eqnarray}
Eqs.~(\ref{eqs2psm}) and (\ref{eqspsifi}) indicate that the Hamiltonian
\begin{equation}
\hat H = \hat S^2 + \sqrt{\hbar\Omega}\,\hat S\,,
\label{eqh}
\end{equation}
where $\Omega$ is a constant, has the eigenstates $\mid
\Psi_m\rangle_\pm$ 
\begin{equation}
\hat H\mid\Psi_m\rangle_\pm = \left(\varepsilon_{m+1}
\pm\sqrt{\hbar\Omega}\sqrt{\varepsilon_{m+1}}\right)\mid\Psi_m\rangle_
\pm \label{eqpsipm}
\end{equation}
with the exception of the ground state. It is easy to show that the
ground state is
\begin{equation}
\mid \Psi_0\rangle  =  \left[ \matrix{ 0 \cr \mid 0 \rangle
\cr}\right], \label{grstate}
\end{equation}
with eigenvalue 0. To
emphasize the structure of Eq. (\ref{eqpsipm}) as the generalized
Jaynes-Cummings Hamiltonian we rewrite it as
\begin{equation}
\hat H = \hat A^\dagger\hat A + {1\over 2}\left[\hat A,\hat
A^\dagger\right]\left(\sigma_3+1\right) +
\sqrt{\hbar\Omega}\left(\sigma_+\hat A+\sigma_-\hat A^\dagger\right)\,.
\label{eqhaa}
\end{equation}

When $\hat A$ describes the annihilation operator for the harmonic
oscillator,  $\left[\hat A,\hat A^\dagger\right] = \hbar\omega$, where
$\omega$ is the oscillator frequency. In this case Eq.~(\ref{eqhaa})
reduces to the standard Jaynes-Cummings Hamiltonian.

When $\hat A^\dagger\hat A$ describes the Morse Hamiltonian,
Eq.~(\ref{eqhaa}) takes the form
\begin{eqnarray}
\hat H &=& {\hat p^2\over 2M}+V_0\left(e^{-2\lambda x}-2e^{-\lambda
x}\right) + \sqrt{V_0}{\hbar\lambda\over\sqrt{2M}}\left(\sigma_3 +
1\right)e^{-\lambda x}\nonumber\\ & & + \sqrt{\hbar\Omega
V_0}\left[\sigma_1 \left(1-{\hbar\lambda\over
2\sqrt{2MV_0}}-e^{-\lambda x}\right)-\sigma_2{\hat
p\over\sqrt{2MV_0}}\right]
\label{eqhmorse}
\end{eqnarray}
with the energy eigenvalues 
\begin{eqnarray}
E_m &=&
\sqrt{V_0}{\hbar\lambda\over\sqrt{2M}}(m+1)\left[2-{\hbar\lambda\over
\sqrt{2MV_0}}(m+2)\right]\nonumber\\ & & \pm
\left\{\hbar\Omega\sqrt{V_0}{\hbar\lambda\over\sqrt{2M}}(m+1)\left[2-
{\hbar\lambda\over\sqrt{2MV_0}}(m+2)\right]\right\}^{1\over 2}\,.
\label{eqemorse}
\end{eqnarray} 

Both harmonic oscillator and Morse potential are shape-invariant
potentials where parameters are related by a translation. It is also
straightforward to use those shape-invariant potentials where the
parameters are related by a scaling \cite{ref19} in writing down
Eq.~(\ref{eqhaa}).

\section{Conclusions}

In this article we introduced a class of shape-invariant bound-state
problems which represent two-level systems. The corresponding
coupled-channel Hamiltonians generalize the Jaynes-Cummings
Hamiltonian. If we take $\hat H_1$ to be the simplest shape-invariant
system, namely the harmonic oscillator, our Hamiltonian,
Eq.~(\ref{eqhaa}), reduces to the standard Jaynes-Cummings
Hamiltonian, which has been extensively used to model a single field
mode on resonance with atomic transitions. 

In this article we only addressed generalization of the
Jaynes-Cummings model to other shape-invariant bound state systems. 
Supersymmetric quantum mechanics has been applied to alpha particle
\cite{ref20} and Coulomb \cite{ref21} scattering problems. More
recently shape-invariance was utilized to calculate quantum tunneling 
probabilities \cite{ref22}. It may be possible to generalize our
results to such continuum problems. Such an investigation will be
deferred to a later publication. 

\section*{ACKNOWLEDGMENTS}

This work was supported in part by the U.S. National Science
Foundation Grant No.\ PHY-9605140 at the University of Wisconsin, and
in part by the University of Wisconsin Research Committee with funds
granted by the Wisconsin Alumni Research Foundation.   A.B.B.\
acknowledges the support of the Alexander von
Humboldt-Stiftung. M.A.C.R.\  acknowledges the support of Funda\c
c\~ao de Amparo \`a Pesquisa do Estado de S\~ao Paulo (Contract No.\
98/13722-2). A.N.F.A.  acknowledges the support of Funda\c c\~ao
Coordena\c c\~ao de Aperfei\c coamento de Pessoal de N\'{\i}vel
Superior (Contract No.  BEX0610/96-8). A.B.B. thanks to the
Max-Planck-Institut f\"ur Kernphysik and M.A.C.R. to the Nuclear
Theory Group at University of Wisconsin for the very kind
hospitality.


\end{document}